\documentclass[journal,letterpaper]{IEEEtran}

\usepackage[dvips]{epsfig}
\usepackage[dvips]{graphicx}
\usepackage{amsmath,amsfonts,bm,amssymb,psfrag,ifthen,color,subfigure}
\usepackage{algpseudocode}
\usepackage{algorithm}
\usepackage{algorithmicx}
\usepackage{setspace}
\usepackage{cite}
\usepackage{longtable}
\usepackage{stfloats}  
\usepackage{graphics,booktabs,color,subfigure}
\usepackage{float}
\usepackage{url}
\usepackage{indentfirst}
\usepackage{tabularx}
\usepackage{multirow}
\usepackage{setspace}

\begin{document}
\title{\LARGE{Robust Beamforming for Localization-Aided Millimeter Wave Communication Systems}}
\author{Junchang Sun, \emph{Student Member, IEEE}, Shuai Ma, \emph{Member, IEEE}, Shiyin Li, Ruixin Yang, Minghui Min, and Gonzalo Seco-Granados, \emph{Senior Member, IEEE}\vspace{-2.5em}
\thanks{The work of J. Sun was supported by the China Scholarship Council (CSC) under Grant 202106420038. The work of  S. Ma  was supported in part by the Open Research Fund of National Mobile Communications Research Laboratory, Southeast University under Grant 2021D02, and in part by the Open Fund of IPOC (BUPT). The work of S. Li was supported by the National Natural Science Foundation of China under Grant 61771474. The work of G. Seco-Granados was supported in part by the Spanish Project under Grant PID2020-118984GB-I00, and in part by the Catalan ICREA Academia Programme. (\emph{Corresponding author: Shiyin Li.})}
\thanks{J. Sun is with the School of Information and Control Engineering, China University of Mining and Technology, Xuzhou 221116, China, also with the Department of Telecommunications and Systems Engineering, Universitat Aut{\`o}noma de Barcelona, 08193 Barcelona, Spain, and also with the Engineering Research Center of Intelligent Control for Underground Space, Ministry of Education, China University of Mining and Technology, Xuzhou, 221116, China (e-mail: sunjc@cumt.edu.cn).}
\thanks{S. Ma is with the School of Information and Control Engineering,  China University of Mining and Technology, Xuzhou 221116, China, also with the National Mobile Communications Research Laboratory, Southeast University, Nanjing 210096, China, and also with the Shanxi Key Laboratory of Information Communication Network and Security, Xian University of Posts and Telecommunications, Xian 710121, China (e-mail: mashuai001@cumt.edu.cn).}
\thanks{S. Li, R. Yang, and M. Min are with the School of Information and Control  Engineering, China
University of Mining and Technology, Xuzhou 221116,
China (e-mail: \{lishiyin, ray.young, minmh\}@cumt.edu.cn).}
 \thanks{G. Seco-Granados is with the Department of Telecommunications and Systems Engineering, Universitat Aut{\`o}noma de Barcelona, 08193 Barcelona, Spain (e-mail: gonzalo.seco@uab.cat).}
 }

\maketitle
\begin{abstract}
In this letter, we investigate a robust beamforming problem for localization-aided millimeter wave (mmWave) communication systems. To handle this problem, we propose a novel restriction and relaxation  (R$\&$R) method. The proposed R$\&$R method aims at minimizing the total transmit power while the positioning error follows a Gaussian distribution. Specifically, in the restriction phase of R$\&$R, the probabilistic constraint is transformed into the deterministic form by using the Bernstein-type inequality. In the relaxation phase of R$\&$R, the non-convex optimization problem is reformulated into a convex semidefinite program (SDP) by using semidefinite relaxation (SDR) and first-order Taylor expansion methods. To the best of our knowledge, we first consider the impact of the distribution of the positioning error on the channel state information (CSI), which further influences the data rate. Numerical results present the trade-off of the beamforming between the communication and positioning.
\end{abstract}\vspace{-1.5em}

\vspace{-1.5em}

\section{Introduction}
Simultaneous precise positioning and high-quality communication are becoming a widespread requirement in the industrial Internet of Things (IoT). 
The conventional GPS is weak or unavailable for communication while ensuring the positioning requirement in harsh environments. 
Generally, the millimeter wave (mmWave) has been shown to be effective in the field of both wireless communication and localization applications.

Most of existing works of location-aided communication systems ignore the impact of the positioning error distribution \cite{Yin2019WCL}. As mentioned in \cite{Yin2019WCL}, a novel positioning-communication integrated signal is designed to achieve a high-accuracy range measurement. In \cite{Jeong2018}, a sum-rate maximization problem under the positioning constraint and a positioning-error minimization problem under the communication constraint are optimized for a massive full-dimensional multi-input multi-output (MIMO) system. 
The beamforming vector is optimized to reduce the localization error bound by using a novel successive localization and beamforming scheme for the 5G mmWave MIMO system in \cite{Zhou2019TSP}. 
In \cite{Jeong2015}, the optimal beamforming design problem is investigated under the independent data rate constraint and positioning constraint. 
However, the coupling relationship between the positioning error distribution and the quality of communications has not been explored yet. 

Differ from existing works, we propose a novel restriction and relaxation  (R$\&$R) method to investigate the robust optimal beamforming problem via considering the distribution of the positioning error. The positioning error influences the estimation of the channel state information (CSI), which in turn affects the data rate. 
Our objective is to minimize the total transmit power under the robust data rate while the positioning error follows a Gaussian distribution. Specifically, the positioning error is evaluated by the Cram\'er-Rao bound (CRB).
To solve this problem, we first take the restriction phase, where the probabilistic constraint is transformed into a deterministic form via using the Bernstein-type inequality. Then, we take the relaxation phase, where the non-convex problem is transformed into a convex problem via using semidefinite relaxation (SDR) and first-order Taylor expansion methods.

{\em Notations:} $a$, ${\bm{a}}$, ${\bm{A}}$, and ${\cal A}$ denote a scalar, vector, matrix, and set, respectively. $\Re \left\{  \cdot  \right\}$ and  ${\Im \left\{  \cdot  \right\}}$ denote real and imaginary parts, respectively. 
${\rm{rank}}\left(  \cdot  \right)$, ${\rm{tr}}\left\{  \cdot  \right\}$, ${\left|  \cdot  \right|}$, ${\left\|  \cdot  \right\|_2}$, ${\left\|  \cdot  \right\|_{\rm{F}}}$, ${\left(  \cdot  \right)^{\rm{T}}}$, ${\left(  \cdot  \right)^{\rm{H}}}$, and ${\left(  \cdot  \right)^{-1}}$ denote rank, trace, absolution, ${\ell _{\rm{2}}}$ norm, Frobenius norm, transpose, complex transpose, and inverse respectively.
${\bm{A}} \succeq {\bm{B}}$ means that matrix ${\bm{A}} - {\bm{B}}$ is positive semidefinite. ${\mathbb E}\left\{  \cdot  \right\}$ and $\Pr \left\{  \cdot  \right\}$ denote the expectation the probability operator, respectively. ${\bm{I}}$ is the identity matrix, ${\cal I} = {\left[ {1,0,0} \right]^{\rm{T}}}$, and ${\cal M} \buildrel \Delta \over = \left\{ {1, \cdots ,M} \right\}$.

\section{System model}
We consider a distribution system containing several multi-antenna base stations (BSs) and a single-antenna user. Each BS is equipped with ${{N_{\rm{B}}}}$ antennas. Let denote the location of the $i$th BS as ${{\bm{p}}_i} \buildrel \Delta \over = {\left[ {{p_{x,i}},{p_{y,i}}} \right]^{\rm{T}}}$, $i \in {\cal M}$, where $M$ denotes the number of BSs. Moreover, the location of the user is unknown, denoted by ${\bm{u}} \buildrel \Delta \over = {\left[ {{u_x}},{u_y} \right]^{\rm{T}}}$.

\subsection{Positioning Frame}
The transmit positioning signal of the $i$th BS ${{\bm{x}}_{{\rm{p}},i}}\left( t \right)$ with the duration ${T_{\rm{p}}}$ is modeled by
\begin{align}
{{\bm{x}}_{{\rm{p}},i}}\left( t \right) = {{\bm{w}}_i}{s_{{\rm{p}},i}}\left( t \right),
\end{align}
where ${\bm{w}}_i$ and ${s_{{\rm{p}},i}}\left( t \right)$ denote the beamforming vector and positioning pilot signal of the $i$th BS, satisfying ${\mathbb E}\left\{ {s_{{\rm{p}},i}^2\left( t \right)} \right\} = 1$.

The received positioning signal of the line-of-sight (LoS) link ${y_{{\rm{p}},i}}\left( t \right)$ is given by 
\begin{align}\label{receive}
{y_{{\rm{p}},i}}\left( t \right) = {\Lambda _i}{s_{{\rm{p}},i}}\left( {t - {\tau _i}} \right) + {n_{{\rm{p}},i}}\left( t \right),
\end{align}
where
\begin{align}\label{Lambda}
{\Lambda _i} = {\bm{g}}_i^{\rm{H}}{{\bm{w}}_i},
\end{align}
${\tau _i} = \frac{{{{\left\| {{\bm{u}} - {{\bm{p}}_i}} \right\|}_2}}}{c} + b$ represents the effective delay between the $i$th BS and user, $b$ denotes the clock bias, which is  modeled as an independent Gaussian random variables with zero mean and variance $\sigma _b^2$, $c$ denotes the transmit speed, and ${n_{{\rm{p}},i}}\left( t \right)$ is the circularly symmetric complex Gaussian (CSCG) noise with zero mean and two-sided power spectral density (PSD) ${N_{\rm{p}}}$ for the positioning signal of the $i$th BS.

Moreover, ${{\bm{g}}_i}$ in \eqref{Lambda} is the complex channel vector between the $i$th BS and the user, which is expressed as
\begin{align}
{{\bm{g}}_i} = \sqrt {{N_{\rm{B}}}} {\rho _i}{h_i}{{\bm{a}}\left( {{\theta _i}} \right)},
\end{align}
where  ${\rho _i}$ and $h _i$ denote the path loss and the complex channel gain of the $i$th BS, respectively, and ${{\theta _i}}$ is the angle of departure (AOD) of the $i$th BS.
Furthermore, we adopt uniform linear array (ULA) as transmit antenna array, so the ${{\bm{a}}\left( \theta_i  \right)}$ is denoted by
\begin{align}
{\bm{a}}\left( {{\theta _i}} \right) = \frac{1}{{\sqrt {{N_{\rm{B}}}} }}{\left[ {1,{e^{j\frac{{2\pi }}{\lambda }\Delta d\sin {\theta _i}}}, \cdots ,{e^{j\left( {{N_{\rm{B}}} - 1} \right)\frac{{2\pi }}{\lambda }\Delta d\sin {\theta _i}}}} \right]^{\rm{T}}},
\end{align}
where $\lambda$ denotes wave length, and $\Delta d = \frac{{{\lambda }}}{2}$ is inter-element distance. For notational convenience, we write ${\bm{a}}\left( \theta _i  \right)$ as ${\bm{a}_i}$ in following contents.

Based on the received signal ${y_{{\rm{p}},i}}\left( t \right)$, we can estimate a rough location ${\bm{\hat u}}$ of the user.
Subsequently, the positioning error $\Delta {\bm{u}}$ is given by
\begin{align}
\Delta {\bm{u}} = {\bm{u}} - {\bm{\hat u}}.
\end{align}

Then, we calculate Fisher information matrix (FIM) and equivalent FIM (EFIM) \cite{Abu-Shaban2020} as follows.
Define the unknown parameter set ${\bm{\eta}}  \buildrel \Delta \over = {\left[ {{\bm{\xi }}_1^{\rm{T}}, \cdots ,{\bm{\xi }}_M^{\rm{T}}} \right]^{\rm{T}}}$, where ${{\bm{\xi }}_i} \buildrel \Delta \over = {\left[ {{\tau _i},{\bm{\Lambda}} _i^{\rm{T}}} \right]^{\rm{T}}}, {{\bm{\Lambda }}_i} = {\left[ {\Re \left\{ {{\Lambda _i}} \right\},\Im \left\{ {{\Lambda _i}} \right\}} \right]^{\rm{T}}}, i \in {\cal M}$. The mean square error (MSE) of unbiased estimation ${{\bm{\hat \eta }}}$ of ${\bm{\eta }}$ is given by
\begin{align}
{\mathbb E}\left\{ {\left( {{\bm{\hat \eta }} - {\bm{\eta }}} \right){{\left( {{\bm{\hat \eta }} - {\bm{\eta }}} \right)}^{\rm{T}}}} \right\} \succeq {\bm{J}}_{\bm{\eta }}^{ - 1},
\end{align}
where ${{\bm{J}}_{\bm{\eta}} }$ is the FIM. Furthermore, defining location related unknown parameter vector ${\bm{\tilde \eta }} \buildrel \Delta \over = {\left[ {{{\bm{u}}^{\rm{T}}},{\bm{\Lambda}} _1^{\rm{T}}, \cdots ,{\bm{\Lambda}} _M^{\rm{T}}},b \right]^{\rm{T}}}$, the EFIM ${\bm{J}}_{\rm{p}}^{\rm{e}}$ of the position parameter is given by
\begin{align}\label{Jpe}
{\bm{J}}_{\rm{p}}^{\rm{e}} =& \Xi \sum\limits_{i \in {\cal M}} {{{\left| {{\Lambda _i}} \right|}^2}{{\bm{\alpha }}_i}{\bm{\alpha }}_i^{\rm{T}}} - \frac{{{\Xi ^2}}}{{\Xi \sum\limits_{i \in {\cal M}} {{{\left| {{\Lambda _i}} \right|}^2}}  + \frac{1}{{\sigma _b^2}}}} \times \qquad \nonumber \\
  &\qquad\qquad\qquad\ \left( {\sum\limits_{i \in {\cal M}} {{{\left| {{\Lambda _i}} \right|}^2}{{\bm{\alpha }}_i}} } \right)\left( {\sum\limits_{i \in {\cal M}} {{{\left| {{\Lambda _i}} \right|}^2}{\bm{\alpha }}_i^{\rm{T}}} } \right),
\end{align}
where
\begin{subequations}
\begin{align}
\Xi &= \frac{{4{\pi ^2}W_{{\rm{eff}}}^2}}{{{N_{\rm{p}}}}},\\
{{\bm{\alpha }}_i} &= {\left[ {\frac{{\partial {\tau _i}}}{{\partial {u_x}}},\frac{{\partial {\tau _i}}}{{\partial {u_y}}}} \right]^{\rm{T}}} = \frac{1}{c}{\left[ {\frac{{{u_x} - {p_{i,x}}}}{{{{\left\| {{\bm{u}} - {{\bm{p}}_i}} \right\|}_2}}},\frac{{{u_y} - {p_{i,y}}}}{{{{\left\| {{\bm{u}} - {{\bm{p}}_i}} \right\|}_2}}}} \right]^{\rm{T}}}.
\end{align}
\end{subequations}
The details are provided in Appendix. 

\subsection{Communication Frame}
Base on estimated distances, we choose the nearest BS to communicate with the user, marked as the ${i^* }$th BS, with coordinate ${\bm p}$.
The transmit communication signal of the ${i^* }$th BS ${\bm{x}}_{\rm{c}}^* \left( t \right)$ with the duration ${T_{\rm{c}}}$ is modeled by
\begin{align}
{\bm{x}}_{\rm{c}}^* \left( t \right) = {\bm{w}}s_{\rm{c}}^* \left( t \right),
\end{align}
where $s_{\rm{c}}^* \left( t \right)$ denotes the communication pilot signal of the ${i^* }$th BS, satisfying ${\mathbb E}\left\{ {s{{_{\rm{c}}^* }^2}\left( t \right)} \right\} = 1$. 

The received communication signal of the ${i^* }$th BS $y_{\rm{c}}^* \left( t \right)$ is given by
\begin{align}
y_{\rm{c}}^* \left( t \right) = {\left( {{{{\bm{\hat g}}}^* } + \Delta {{\bm{g}}^* }} \right)^{\rm{H}}}{\bm{w}}s_{\rm{c}}^* \left( {t - \tau } \right) + n_{\rm{c}}^* \left( t \right),
\end{align}
where $n_{\rm{c}}^* \left( t \right)$ is the CSCG noise with zero mean and two-sided PSD ${N_{\rm{c}}}$ for the communication signal of the BS, and ${{{{\bm{\hat g}}}^* }}$ and ${\Delta {{\bm{g}}^* }}$ represent the estimated channel and channel error of the ${i^* }$th BS, respectively. 

In order to mathematically formulate the true data rate of the communication frame, we split the true channel into an estimate and an error value. Based on the estimated location ${\bm{\hat u}}$, location error $\Delta {\bm{u}}$, and ${\rho } = \frac{\lambda }{{4\pi {{\left\| {{\bm{ u}} - {{\bm{p}}}} \right\|}_2}}}$, the estimated channel ${{{{\bm{\hat g}}}^* }}$ and channel error ${\Delta {{\bm{g}}^* }}$ can be expressed as
\begin{align}\label{channel}
{{{{\bm{\hat g}}}^* }} &= \frac{{\sqrt {{N_{\rm{B}}}} \lambda {h} }}{{4\pi }}\frac{1}{{{{\left\| {{\bm{\hat u}} - {{\bm{p}}}} \right\|}_2}}}{\bm{\hat a}},\\
{\Delta {{\bm{g}}^* }} &= \frac{{\sqrt {{N_{\rm{B}}}} \lambda {h}}}{{4\pi }}\left( {\frac{1}{{{{\left\| {{\bm{\hat u}} + \Delta {\bm{u}} - {{\bm{p}}}} \right\|}_2}}}{\bm{a}} - \frac{1}{{{{\left\| {{\bm{\hat u}} - {{\bm{p}}}} \right\|}_2}}}{\bm{\hat a}}} \right),\label{channel_2}
\end{align}
where ${ \theta } \buildrel \Delta \over = {\hat \theta } + \Delta {\theta }$ in ${\bm a }$,
${{\hat \theta }} =  {\sin ^{ - 1}}\left( {\frac{{{{\left( {{\bm{\hat u}} - {{\bm{p}}}} \right)}^{\rm{T}}}{{\bm{e}}_y}}}{{{{\left\| {{\bm{\hat u}} - {{\bm{p}}}} \right\|}_2}}}} \right)$ in ${\bm{\hat a}}$ denotes the estimated angle, $\Delta {\theta }$ denotes the angle error, and ${{\bm{e}}_y} \buildrel \Delta \over = {\left[ {0,1} \right]^{\rm{T}}}$ is unit orientation vector. Due to the case of that ${\left\| {{\bm{\hat u}} - {\bm{p}}} \right\|_2} \gg {\left\| {\Delta {\bm{u}}} \right\|_2}$, the effect of the angle error is negligible, i.e., $\Delta {\theta } = 0$. Subsequently, \eqref{channel_2} is transformed into
\begin{align}\label{channel_error}
{\Delta {{\bm{g}}^* }} = \frac{{\sqrt {{N_{\rm{B}}}} \lambda {h} }}{{4\pi }}{\bm{\hat a}}\left( {\frac{1}{{{{\left\| {{\bm{\hat u}} + \Delta {\bm{u}} - {{\bm{p}}}} \right\|}_2}}} - \frac{1}{{{{\left\| {{\bm{\hat u}} - {{\bm{p}}}} \right\|}_2}}}} \right).
\end{align}

Therefore, the data rate (bps/Hz) of the ${i^* }$th BS ${R^*}$ is given by
\begin{align}\label{rate}
{R^*} = \frac{{{T_{\rm{c}}}}}{{{T_{\rm{p}}} + {T_{\rm{c}}}}}{\log _2}\left( {1 + \frac{{{{\left| {{{\left( {{{{\bm{\hat g}}}^* } + {\Delta {{\bm{g}}^* }}} \right)}^{\rm{H}}}{{\bm{w}}}} \right|}^2}}}{{{N_{\rm{c}}}}}} \right).
\end{align}

\section{Problem formulation}
We consider the following optimization problem:

{\bf{Robust Beamforming Problem}}: We denote the rate threshold $\bar R$ of the user and tolerable outage probability ${P_{\rm{out}}}$, solve
\begin{subequations}\label{ori_problem}
\begin{align}
\mathop {\min }\limits_{\bm{w}} &\  {\left\| {{{\bm{w}}}} \right\|_2^2} \\ \label{ori_problem_2}
{\rm{s}}.{\rm{t}}.&\ \Pr \left\{ {{R^*} \le \bar R} \right\} \le {P_{\rm{out}}}.
\end{align}
\end{subequations}

By substituting \eqref{channel} and \eqref{channel_error} into \eqref{rate}, we can write the constraint \eqref{ori_problem_2} as
\begin{align}\label{eq17}
\Pr \left\{ {\left\| {{\bm{\hat u}} + \Delta {\bm{u}} - {{\bm{p}}}} \right\|_2^2 \ge \gamma {{\bm{\hat a}}^{\rm{H}}}{{{\bm{w}}}{\bm{w}}^{\rm{H}}}{\bm{\hat a}}} \right\} \le {P_{\rm{out}}},
\end{align}
where $\gamma {\rm{ = }}\frac{{{\lambda ^{\rm{2}}}{N_{\rm{B}}}{{\left| {h} \right|}^2}}}{{{{\left( {{\rm{4}}\pi } \right)}^{\rm{2}}}{N_{\rm{c}}}\left( {{2^{\frac{{{T_{\rm{p}}} + {T_{\rm{c}}}}}{{{T_{\rm{c}}}}}\bar R}} - 1} \right)}}$.

The optimization problem \eqref{ori_problem} is hard to be solved due to non-convex objective function and constraint \eqref{eq17}. To handle this problem, we propose a novel R\&R method, which consists of a restriction phase and a relaxation phase.

\subsection{Restriction Phase}
Define ${{\bm{\Sigma }}} \buildrel \Delta \over = {{\bm{w}}}{\bm{w}}^{\rm{H}}, {{\bm{V}}} \buildrel \Delta \over = {{\bm{\hat a}}}{\bm{{\hat a}}^{\rm{H}}}$, the optimization problem \eqref{ori_problem} can be reformulated as
\begin{subequations}\label{problem}
\begin{align}
\mathop {\min }\limits_{\bm{\Sigma }} & \ {{\rm{tr}}\left\{ {{{\bm{\Sigma }}}} \right\}} \\ \label{problem_2}
{\rm{s}}.{\rm{t}}.&\ \Pr \left\{ {\left\| {{\bm{\hat u}} + \Delta {\bm{u}} - {{\bm{p}}}} \right\|_2^2 \ge \gamma {\rm{tr}}\left\{ {{{\bm{\Sigma }}}{{\bm{V}}}} \right\}} \right\} \le {P_{\rm{out}}},\\
&\ {\rm{rank}}\left( {{{\bm{\Sigma }}}} \right) = 1\\
&\ {{\bm{\Sigma }}} \succeq {\bm{0}}.
\end{align}
\end{subequations}
Note that the probabilistic constraint \eqref{problem_2} does not have a closed form.
To track this challenging problem, we assume that the positioning error ${\Delta {\bm{u}}}$ follows a Gaussian distribution \cite{Wang2009TSP}. 
Equivalently, the ${\Delta {\bm{u}}}$ can be written as $\Delta {\bm{u}} = {\left( {{\bm{J}}_{\rm{p}}^{\rm{e}}} \right)^{ - \frac{1}{2}}}{{\bm{e}}_{\rm{p}}}$, where ${{\bm{e}}_{\rm{p}}} \sim {\cal N}\left( {{\bm{0}},{\bm{I}}} \right)$.
Subsequently, the constraint \eqref{problem_2} can be written as
\begin{align}\label{probability}
\Pr \left\{ {{\bm{e}}_{\rm{p}}^{\rm{T}}{{\left( {{\bm{J}}_{\rm{p}}^{\rm{e}}} \right)^{ - 1}}}{{\bm{e}}_{\rm{p}}} + 2{\bm{e}}_{\rm{p}}^{\rm{T}}{{\bm{r}}} \ge {\nu }} \right\} \le {P_{\rm{out}}},
\end{align}
where ${\bm{r}} \buildrel \Delta \over = {\left( {{\bm{J}}_{\rm{p}}^{\rm{e}}} \right)^{ - \frac{1}{2}}}\left( {{\bm{\hat u}} - {{\bm{p}}}} \right)$, and ${\nu } \buildrel \Delta \over = \gamma {\rm{tr}}\left\{ {{{\bm{\Sigma }}}{{\bm{V}}}} \right\} - \left\| {{\bm{\hat u}} - {{\bm{p}}}} \right\|_2^2$.
Then, we use the following lemma to transform the probabilistic constraint \eqref{probability} to the deterministic form.

{\bf{Lemma 1 (Bernstein-type inequality)}}\cite{Wang2014}: Let $\chi  = {{\bm{z}}^{\rm{T}}}{\bm{Sz}} + 2\Re \left\{ {{{\bm{z}}^{\rm{T}}}{\bm{s}}} \right\}$, where ${\bm{S}} \in {{\mathbb H}^N}$ denotes a complex hermitian matrix, ${{\bm{s}} \in {{\mathbb R}^N}}$, and ${\bm{z}} \sim {\cal N}\left( {{\bm{0}},{\bm{I}}} \right)$. Then, for the any given constant $\zeta  > 0$, we have 
\begin{align}
\begin{small}
\Pr \left\{ {\chi  \ge {\rm{tr}}\left\{ {\bm{S}} \right\} + \sqrt {2\zeta } \sqrt {\left\| {\bm{S}} \right\|_{\rm{F}}^2 + 2\left\| {\bm{s}} \right\|_2^2}  + \zeta {{\lambda}^ + }\left( {\bm{S}} \right)} \right\} \le {e^{ - \zeta }},
\end{small}
\end{align}
where ${{\lambda ^ + }\left( {\bm{S}} \right) = \max \left\{ {{\lambda _{\max }}\left( {\bm{S}} \right),0} \right\}}$ and ${{\lambda _{\max }}\left( {\bm{S}} \right)}$ is the maximum eigenvalue of the matrix ${\bm{S}}$.

By using the Bernstein-type inequality in lemma 1, the probabilistic constraint \eqref{probability} can be reformulated as
\begin{small}
\begin{align}\label{deter_const}
{{\rm{tr}}\left\{ {{\left( {{\bm{J}}_{\rm{p}}^{\rm{e}}} \right)^{ - 1}}} \right\} + \sqrt {2\zeta } \sqrt {\left\| {{\left( {{\bm{J}}_{\rm{p}}^{\rm{e}}} \right)^{ - 1}}} \right\|_{\rm{F}}^2 + 2\left\| {{{\bm{r}}}} \right\|_2^2}  + \zeta {\lambda ^ + } {{\left( {{\bm{J}}_{\rm{p}}^{\rm{e}}} \right)^{ - 1}}} \le {\nu }},
\end{align}
\end{small}
where $\zeta  =  - \ln \left( {{P_{out}}} \right)$. Equivalently, the constraint \eqref{deter_const} can be written as
\begin{subequations}
\begin{align}
{\rm{tr}}\left\{ {{\left( {{\bm{J}}_{\rm{p}}^{\rm{e}}} \right)^{ - 1}}} \right\} + \sqrt {2\zeta } \varpi  + \zeta \varrho - {\nu } &\le 0, \label{fnorm_1} \\
\left\| {{\left( {{\bm{J}}_{\rm{p}}^{\rm{e}}} \right)^{ - 1}}} \right\|_{\rm{F}}^2 + 2\left\| {{{\bm{r}}}} \right\|_2^2\le \varpi ^2, \varpi  &\ge 0, \label{fnorm_2} \\
{\varrho}{\bm{I}} - {{\left( {{\bm{J}}_{\rm{p}}^{\rm{e}}} \right)^{ - 1}}}\succeq {\bm 0},{\varrho} &\ge 0, \label{fnorm_3}
\end{align}
\end{subequations}
where $\varpi$ and ${\varrho}$ are auxiliary variables.

Consequently, the robust beamforming problem under the Gaussian assumption is reformulated as
\begin{subequations}\label{Gaussian_problem}
\begin{align}
\mathop {\min }\limits_{{\bm{\Sigma }},{\varrho },{\varpi }} &\ {{\rm{tr}}\left\{ {{{\bm{\Sigma }}}} \right\}}\\
{\rm{s}}.{\rm{t}}. &\ {\rm{rank}}\left( {{{\bm{\Sigma }}}} \right) = 1, \label{Gaussian_problem_1}\\
&\ {{\bm{\Sigma }}} \succeq {\bm{0}},\\
&\ \eqref{fnorm_1}, \eqref{fnorm_2}, \eqref{fnorm_3}. \nonumber
\end{align}
\end{subequations}

However, the solution is still complicated by the presence of non-convex constraints \eqref{Gaussian_problem_1}, \eqref{fnorm_2}, and \eqref{fnorm_3}. Specifically, the constraint \eqref{Gaussian_problem_1} is non-convex caused by the rank operate.
The constraints \eqref{fnorm_2} and \eqref{fnorm_3} are non-convex by the presence of the inverse matrix and quadratic $\varpi ^2$, which is nonlinear and non-convexity-preserving. Therefore, we use the relaxation method to handle it.

\subsection{Relaxation Phase}
For the constraint \eqref{Gaussian_problem_1}, we can solve it by using the semidefinite relaxation (SDR) method \cite{Luo2010}.
For constraints \eqref{fnorm_2} and \eqref{fnorm_3}, we can convert these into convex approximate forms by using the first-order Taylor expansion. Specifically,  the $\varpi ^2$, ${\left( {{\bm{J}}_{\rm{p}}^{\rm{e}}} \right)^{ - 1}}$, $\left\| {{{\bm{r}}}} \right\|_2^2$, and $\left\| {{\left( {{\bm{J}}_{\rm{p}}^{\rm{e}}} \right)^{ - 1}}} \right\|_{\rm{F}}^2$ can be approximated as
\begin{small}
\begin{subequations}\label{Taylor}
\begin{align}
&\varpi ^2\approx \varpi _{0}^2 + 2{\varpi _{0}}\left( {{\varpi } - {\varpi _{0}}} \right),\\
&{{{\left( {{\bm{J}}_{\rm{p}}^{\rm{e}}} \right)^{ - 1}}}}\approx {\left( {{\bm{J}}_{{\rm{p}},0}^{\rm{e}}} \right)^{ - 1}} - {\left( {{\bm{J}}_{{\rm{p}},0}^{\rm{e}}} \right)^{ - 2}}\left( {{\bm{J}}_{{\rm{p}}}^{\rm{e}} - {\bm{J}}_{{\rm{p}},0}^{\rm{e}}} \right),\\
&\left\| {{{\bm{r}}}} \right\|_2^2 \approx {\left( {{\bm{\hat u}} - {{\bm{p}}}} \right)^{\rm{T}}}{\left( {\left( {{\bm{J}}_{{\rm{p}},0}^{\rm{e}}} \right)^{ - 1}} - {\left( {{\bm{J}}_{{\rm{p}},0}^{\rm{e}}} \right)^{ - 2}}\left( {{\bm{J}}_{{\rm{p}}}^{\rm{e}} - {\bm{J}}_{{\rm{p}},0}^{\rm{e}}} \right)\right)}\left( {{\bm{\hat u}} - {{\bm{p}}}} \right),\\
&\left\| {{{\left( {{\bm{J}}_{\rm{p}}^{\rm{e}}} \right)^{ - 1}}}} \right\|_{\rm{F}}^2\approx {\rm{tr}}\left\{ {{{\left( {{\bm{J}}_{{\rm{p}},0}^{\rm{e}}} \right)}^{ - 2}}} \right\} - 2{\rm{tr}}\left\{ {{{\left( {{\bm J}_{{\rm{p}},0}^{\rm{e}}} \right)}^{ - 3}}\left( {{\bm{J}}_{{\rm{p}}}^{\rm{e}} - {\bm{J}}_{{\rm{p}},0}^{\rm{e}}} \right)} \right\},
\end{align}
\end{subequations}
\end{small}
where ${\varpi _{0}} \ge 0$ is a preset constant and ${{\bm{J}}_{{\rm{p}},0}^{\rm{e}}}$ is calculated with any  given preset matrix ${\bm{\Sigma }}_{i,0} \succeq {\bm 0}$ and angle ${\theta _{i,0}}$ for $i \in {\cal M}$. However, the ${\bm{J}}_{\rm{p}}^{\rm{e}}$ is not linear over ${{\bm{\Sigma }}}$, which causes constraints of ${{\bm{\Sigma }}}$ being non-convex. To handle this problem, we use the Remark 1 to scale ${\bm{J}}_{\rm{p}}^{\rm{e}}$ to a linear equation.

{\bf{Remark 1}}\cite{Jeong2015}: Suppose $\frac{1}{{\sigma _b^2}} \gg \Xi \sum\limits_{i \in {\cal M}} {{{\left| {{\Lambda _i}} \right|}^2}} $ in \eqref{Jpe}, we can observe that the following inequality holds:
\begin{align}
{\bm{J}}_{\rm{p}}^{\rm{e}} \succeq {\bm{\tilde J}}_{\rm{p}}^{\rm{e}},
\end{align}
where
\begin{small}
\begin{align}\label{Jpe_hat}
{\bm{\tilde J}}_{\rm{p}}^{\rm{e}} = \Xi \sum\limits_{i \in {\cal M}} {{{{{\left| {{\Lambda _i}} \right|}^2}}}{{\bm{\alpha }}_i}{\bm{\alpha }}_i^{\rm{T}}} - {\left( {\Xi {\sigma _b}} \right)^2}\left( {\sum\limits_{i \in {\cal M}} {{{\left| {{{\bm{g}}_i}} \right|}^2}{{\bm{\alpha }}_i}} } \right)\left( {\sum\limits_{i \in {\cal M}} {{{\left| {{{\bm{g}}_i}} \right|}^2}{\bm{\alpha }}_i^{\rm{T}}} } \right).
\end{align}
\end{small}

Then, we replace ${\bm{J}}_{\rm{p}}^{\rm{e}}$ by ${\bm{\tilde J}}_{\rm{p}}^{\rm{e}}$ 
in \eqref{fnorm_1}, we can obtain
\begin{align}\label{tilde_fnorm_1}
{\rm{tr}}\left\{ {{\left( {{\bm{\tilde J}}_{\rm{p}}^{\rm{e}}} \right)^{ - 1}}} \right\} + \sqrt {2\zeta } \varpi  + \zeta \varrho - {\nu } &\le 0.
\end{align}
Replacing ${\bm{J}}_{\rm{p}}^{\rm{e}}$ by ${\bm{\tilde J}}_{\rm{p}}^{\rm{e}}$ in \eqref{Taylor}, and substitute \eqref{Taylor} into \eqref{fnorm_2} and \eqref{fnorm_3}, we can obtain
\begin{subequations}
	\begin{align}\label{re_constraints_1}
	&{\rm{tr}}\left\{ {{{\left( {{\bm{\tilde J}}_{{\rm{p}},0}^{\rm{e}}} \right)}^{ - 2}}} \right\} - 2{\rm{tr}}\left\{ {{{\left( {{\bm{\tilde J}}_{{\rm{p}},0}^{\rm{e}}} \right)}^{ - 3}}\left( {{\bm{\tilde J}}_{\rm{p}}^{\rm{e}} - {\bm{\tilde J}}_{{\rm{p}},0}^{\rm{e}}} \right)} \right\} + 2{\left( {{\bm{\hat u}} - {{\bm{p}}}} \right)^{\rm{T}}}\nonumber \\
	&\times \left( {{{\left( {{\bm{\tilde J}}_{{\rm{p,0}}}^{\rm{e}}} \right)}^{ - 1}} - {{\left( {{\bm{\tilde J}}_{{\rm{p}},{\rm{0}}}^{\rm{e}}} \right)}^{ - 2}}\left( {{\bm{\tilde J}}_{\rm{p}}^{\rm{e}} - {\bm{\tilde J}}_{{\rm{p,0}}}^{\rm{e}}} \right)} \right) \left( {{\bm{\hat u}} - {{\bm{p}}}} \right)\nonumber \\
	&\le \varpi _{0}^2 + 2{\varpi _{0}}\left( {{\varpi } - {\varpi _{0}}} \right), \varpi \ge 0.\\
	&{\varrho }{\bm{I}} - {\left( {{\bm{\tilde J}}_{{\rm{p}},0}^{\rm{e}}} \right)^{ - 1}} + {\left( {{\bm{\tilde J}}_{{\rm{p}},0}^{\rm{e}}} \right)^{ - 2}}\left( {{\bm{\tilde J}}_{\rm{p}}^{\rm{e}} - {\bm{\tilde J}}_{{\rm{p}},0}^{\rm{e}}} \right) \succeq {\bm{0}},{\varrho } \ge 0. \label{re_constraints_2}
	\end{align}
\end{subequations}

Consequently, the original optimization problem is reformulated as
\begin{subequations}\label{final_problem}
	\begin{align}
	\mathop {\min }\limits_{{\bm{\Sigma }},{\varrho },{\varpi }} &\ {{\rm{tr}}\left\{ {{{\bm{\Sigma }}}} \right\}}\\
	{\rm{s}}.{\rm{t}}.&\ {{\bm{\Sigma }}} \succeq {\bm{0}},\\
	&\ \eqref{tilde_fnorm_1}, \eqref{re_constraints_1}, \eqref{re_constraints_2}. \nonumber
	\end{align}
\end{subequations}

The R\&R method for solving the optimization problem  is detailed in Algorithm 1. We can obtain the optimal matrix ${\bm{\Sigma}}^{{\rm{opt}}}$ of \eqref{final_problem} via an interior point method for the given initialization value ${{\bm{\Sigma }}_{0}}$, by using off-the-shelf convex optimization solvers such as CVX \cite{Grant2014}. Next, we use the eigenvalue decomposition (EVD) method to obtain the optimal solution (if ${\rm{rank}}\left( {{{\bm{\Sigma }}^{{\rm{opt}}}}} \right) = 1$) or the approximate solution (if ${\rm{rank}}\left( {{{\bm{\Sigma }}^{{\rm{opt}}}}} \right) \ne 1$), which is given by
\begin{align}\label{rank-one-app}
	{\bm{\hat w}} = \sqrt {{\lambda _{\max }}\left( {{{\bm{\Sigma }}^{{\rm{opt}}}}} \right)} {\bm{v}}\left( {{{\bm{\Sigma }}^{{\rm{opt}}}}} \right),
\end{align}
where ${\bm{v}}\left( {{{\bm{\Sigma }}^{{\rm{opt}}}}} \right)$ is the eigenvector corresponding to the maximum eigenvalue ${{\lambda _{\max }}\left( {{{\bm{\Sigma }}^{{\rm{opt}}}}} \right)}$. When the acquired approximate solution ${\bm{\hat w}}$ is not feasible for the problem, we rescale it by ${\bm{\hat w}} \leftarrow \left( 1 + {\delta_{\rm{inc}}}\right){\bm{\hat w}}$ for any given positive integer $\delta_{\rm{inc}}$ \cite{Jeong2015}.

\begin{algorithm}[htb]
  \caption{: R\&R method for solving the robust beamforming problem}
  \label{alg1:Alternate iteration}
  \begin{algorithmic}[1]
    \State {\bf{Input:}} $\bar R > 0$, ${P_{\rm{out}}} > 0$, termination parameter $\epsilon \ge 0$, positive integer $\delta_{\rm{inc}}$
    \State {\bf{Initialization:}} ${\bm{\Sigma }}_{i,0} \succeq {\bm 0}$, ${\theta _{i,0}}$,  ${\bm{\tilde J}}_{{\rm{p}},0}^{\rm{e}}$ by \eqref{Jpe_hat}, ${\varpi }_0 > 0$,  and set $n=1$;
    \State Calculate an initial position by ${\bm{\hat u}} = {\bm{u}} + {\left( {\bm{\tilde J}}_{{\rm{p}},0}^{\rm{e}} \right)^{ - \frac{1}{2}}}{{\bm{e}}_{\rm{p}}}$ and angle by ${{\hat \theta }} =  {\sin ^{ - 1}}\left( {\frac{{{{\left( {{\bm{\hat u}} - {{\bm{p}}}} \right)}^{\rm{T}}}{{\bm{e}}_y}}}{{{{\left\| {{\bm{\hat u}} - {{\bm{p}}}} \right\|}_2}}}} \right)$;
    \Repeat
    \State  Obtain ${\bm{\Sigma }}_{n  }$ and ${\varpi }_n$ as solutions of \eqref{final_problem};
    \State Compute ${{\bm{\tilde J}}_{{\rm{p}},n}^{\rm{e}}}$  by using \eqref{Jpe_hat};
    \State Update ${\bm{\tilde J}}_{{\rm{p}},0}^{\rm{e}} \leftarrow {\bm{\tilde J}}_{{\rm{p}},n}^{\rm{e}} $, ${\varpi }_0 \leftarrow {\varpi }_n$, and $n=n+1$;
    \Until $\left| {{\rm{tr}}\left\{ {{{\bm{\Sigma }}_{n}}} \right\} - {\rm{tr}}\left\{ {{{\bm{\Sigma }}_{n-1}}} \right\}} \right| \le \epsilon$
    \State Extract ${\bm{\hat w}}$ by \eqref{rank-one-app};
    \State Check whether the solution ${\bm{\hat w}}$ is feasible or not. If so, return ${\bm{\hat w}}$; otherwise, rescale ${\bm{\hat w}} \leftarrow \left( 1 + {\delta_{\rm{inc}}}\right){\bm{\hat w}}$ until ${\bm{\hat w}}$ is feasible for the problem  and return it;
	\State {\bf{Output:}} Beamforming vector ${\bm{\hat w}}$
  \end{algorithmic}
\end{algorithm}

\section{Numerical Results}
We conduct simulations to evaluate the performance of the localization-aided communication system. The positions of BSs are set as $ {{\left[ {50,50} \right]}^{\rm{T}}}$, $ {{\left[ {75,50} \right]}^{\rm{T}}}$, $ {{\left[ {100,50} \right]}^{\rm{T}}}$, ${{\left[ {50,75} \right]}^{\rm{T}}}$, ${{\left[ {100,75} \right]}^{\rm{T}}}$, $ {{\left[ {50,100} \right]}^{\rm{T}}}$, $ {{\left[ {75,100} \right]}^{\rm{T}}}$, and ${{\left[ {100,100} \right]}^{\rm{T}}}$ in meters. The position of the user is set as ${{\left[ {75,75} \right]}^{\rm{T}}}$ in meters. The other key parameters are listed in Table \ref{parameter}.
\begin{table}[H]
\centering
\small
\caption{Parameters of simulations}
\begin{tabular}{|l|l|}
\hline
Frequency & 60 ${\rm{GHz}}$ \\
\hline
 Outage probability ${P_{\rm{out}}}$ & 0.05 \\
 \hline
 Rate threshold $\bar R$ & 0.3 ${\rm{bps}}/{\rm{Hz}}$\\
 \hline
 Variance of clock bias $\sigma _b$ & 0.01 ${\rm{ns}}$\\
 \hline
 Complex channel gain $\left\{ {{h_i}} \right\}_{i = 1}^M$& $\left( {1 + j} \right)/\sqrt 2 $\\
 \hline
 Two-sided PSD of noise ${N_{\rm{p}}}$, ${N_{\rm{c}}}$ & 1 ${\rm{W}}/{\rm{GHz}}$\\
 \hline
 Effective bandwidth ${W_{{\rm{eff}}}}$ & 125 ${\rm{MHz}}$\\
 \hline
\end{tabular}\label{parameter}
\end{table}
\begin{figure}[htbp]
	\centering
	\includegraphics[width=6cm]{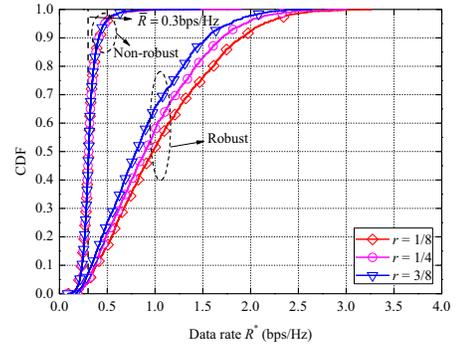}
	\caption{The CDF of the data rate with ${N_{\rm{B}}} = 64$.}\label{cdf}
\end{figure} 
\begin{figure}[htbp]
	\centering
	\includegraphics[width=6.5cm]{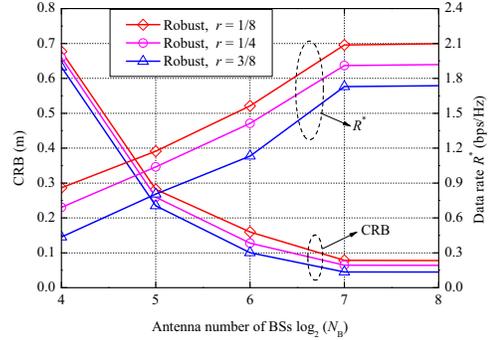}
	\caption{Trade-off between the CRB and the data rate as a function of the antenna number $N_{\rm{B}}$ of BSs.}\label{CRB}
\end{figure}

Fig. \ref{cdf} compares the cumulative distribution function (CDF) of the data rate with different ratios $r = {T_{\rm{p}}}/{T_{\rm{c}}}$, based on 2000 Monte Carlo trials. In each trial, we generate a random estimation location of the user based on any given initial feasible beamforming vectors and AoD. The non-robust condition is also presented for comparison, which takes the term ${{{{\bm{\hat g}}}^* }}$ as the perfect CSI without considering the term ${\Delta {{\bm{g}}^* }}$. The results show the higher data rates for the proposed robust condition compared with the non-robust condition. This suggests that we can achieve the better communication performance by considering the effect of positioning errors. For the robust condition, the data rate is lower with a higher $r$. However, the rate probability $\Pr \left\{ {{R^* } \le \bar R} \right\} > 0.05$ for $r = {\rm{3/8}}$, which slightly violates the rate threshold. The reason is that the less power is allocated to the communication frame.

Fig. \ref{CRB} shows the trade-off between the CRB and data rate when varying the antenna number $N_{\rm{B}}$ of BSs, in which ${\rm{CRB}} = {\rm{tr}}\left\{ {{{\left( {{\bm{\tilde J}}_{\rm{p}}^{\rm{e}}} \right)}^{ - 1}}} \right\}$. For the given $N_{\rm{B}}$, both the CRB and data rate are lower while the more power is allocated to the positioning frame. Moreover, for the given $r$, we can conclude that increasing the $N_{\rm{B}}$ improves the performance of both the data rate and the CRB. The reason is that  the larger number of degrees of freedom is available. Thus, there is no conflict between the performance of these two metrics while increasing the antenna number of BSs.

\section{Conclusion}
In this letter, we have investigated the robust beamforming problem for the localization-aided mmWave communication system. The optimization problem is formulated via considering the distribution of the positioning error. The formulated problem is solved by the proposed R\&R method, which leverages techniques such as Bernstein-type inequality, SDR, and first-order Taylor expansion. Simulation results evaluate the trade-off of the beamforming between the communication and positioning via considering the coupling relationship of the data rate and positioning error distribution.

\begin{appendix}
\section{Derivation of elements in FIM and EFIM}
Denote the received positioning signal vector ${{\bm{y}}_{\rm{p}}}\left( t \right) \buildrel \Delta \over = {\left[ {{y_{{\rm{p}},1}}\left( t \right), \cdots ,{y_{{\rm{p}},M}}\left( t \right)} \right]^{\rm{T}}}$,
the FIM ${{\bm{J}}_{\bm{\eta}} }$ is given by
\begin{align}
{\left[ {{{\bm{J}}_{\bm{\eta}} }} \right]_{{k_1},{k_2}}} = \frac{1}{{{N_{\rm{p}}}}}\int_0^{{T_{\rm{p}}}} {\Re \left\{ {\frac{{\partial {{\bm{\mu }}^{\rm{H}}}\left( t \right)}}{{\partial {k_1}}}\frac{{\partial {\bm{\mu }}\left( t \right)}}{{\partial {k_2}}}} \right\}dt},
\end{align}
where ${k_1},{k_2} \in {\bm{\eta}} $, and
\begin{align}
{\bm{\mu }}\left( t \right) \buildrel \Delta \over = \left[ {\begin{array}{*{20}{c}}
{{\Lambda _1}{s_{{\rm{p}},1}}\left( {t - {\tau _1}} \right)}\\
 \vdots \\
{{\Lambda _M}{s_{{\rm{p}},M}}\left( {t - {\tau _M}} \right)}
\end{array}} \right].
\end{align}

Then, we can write ${{\bm{J}}_{\bm{\eta}} }$ as
\begin{align}
{{\bm{J}}_{\bm{\eta }}} \buildrel \Delta \over = {\rm{diag}}\left\{ {{{\bm{\Phi }}_1}, \cdots ,{{\bm{\Phi }}_M}} \right\},
\end{align}
where
\begin{align}\label{eq33}
{{\bm{\Phi }}_i} &= \left[ {\begin{array}{*{20}{c}}
{\frac{{4{\pi ^2}{{\left| {{\Lambda _i}} \right|}^2}W_{{\rm{eff}}}^2}}{{{N_{\rm{p}}}}}}&{{{\bm{0}}_2}}\\
{{\bm{0}}_2^{\rm{T}}}&{\frac{1}{{{N_{\rm{p}}}}}{{\bm{I}}_2}}
\end{array}} \right],i \in {\cal M}.
\end{align}
In \eqref{eq33}, $W_{{\rm{eff}}}$ denotes the effective bandwidth, which is given by
\begin{small}
\begin{align}
W_{{\rm{eff}}}^2 \buildrel \Delta \over = \int_{ - \infty }^{ + \infty } {{{\left| {fS\left( f \right)} \right|}^2}df}  = \int_{ 0 }^{ T_{\rm{p}} } {\frac{{\partial {s_{{\rm{p}},i}}\left( {t - {\tau _i}} \right)}}{{\partial {\tau _i}}}\frac{{\partial {s_{{\rm{p}},i}}\left( {t - {\tau _i}} \right)}}{{\partial {\tau _i}}}dt},
\end{align}
\end{small}
where ${S\left( f \right)} $ denotes as the Fourier transform of ${s_{{\rm{p}},i}}\left( t \right)$.

Moreover,  given ${\bm{\Upsilon}}  = \frac{{\partial {\bm{\eta }}}}{{\partial {\bm{\tilde \eta }}}}$, the EFIM ${\bm{J}}_{\bm{\tilde \eta }} $ can be expressed as \cite{Jeong2015}
\begin{align}
{{\bm{J}}_{\bm{\tilde \eta }}} = {\bm{\Upsilon}} {{\bm{J}}_{{\bm{ \eta }}}}{{\bm{\Upsilon}} ^{\rm{T}}} + {{\bm{J}}_b},
\end{align}
where
\begin{subequations}
\begin{align}
{\bm{\Upsilon}}  &= \left[ {\begin{array}{*{20}{c}}
{{{\bm{D}}_1}}& \cdots &{{{\bm{D}}_M}}\\
{{{\bm{T}}_1}}&{}&{}\\
{}& \ddots &{}\\
{}&{}&{{{\bm{T}}_M}}\\
{{{\cal I}^{\rm{T}}}}& \cdots &{{{\cal I}^{\rm{T}}}}
\end{array}} \right],\\
{{\bm{D}}_i} &= \left[ {\frac{{\partial {\tau _i}}}{{\partial {\bm{u}}}},{{\bm{0}}_{2 \times 2}}} \right],\\
{{\bm{T}}_i} &= \left[ {\begin{array}{*{20}{c}}
0&1&0\\
0&0&1
\end{array}} \right],\\
{\left[ {{{\bm{J}}_b}} \right]_{m,n}} &= \left\{ \begin{array}{l}
\frac{1}{{\sigma _b^2}},m = n = 2M + 3,\\
\ 0,\ {\rm{otherwise}}.
\end{array} \right.
\end{align}
\end{subequations}
Then, we can express the ${\bm{J}}_{\bm{\tilde \eta }} $ as
\begin{align}
{{\bm{J}}_{{\bm{\tilde \eta }}}} = \left[ {\begin{array}{*{20}{c}}
{\bm{A}}&{\bm{B}}\\
{{{\bm{B}}^{\rm{T}}}}&{\bm{C}}
\end{array}} \right],
\end{align}
where
\begin{subequations}
\begin{align}
{\bm{A}} &= \sum\limits_{i \in {\cal M}} {{{\bm{D}}_i}{{\bm{\Phi }}_i}{\bm{D}}_i^{\rm{T}}},\\
{\bm{B}} &= \left[ {{{\bm{0}}_{2 \times 2M}},\sum\limits_{i \in {\cal M}} {{{\bm{D}}_i}{{\bm{\Phi }}_i}{\cal I}} } \right],\\
{\bm{C}} &= {\rm{diag}}\left\{ {\frac{1}{{{N_{\rm{p}}}}}{{\bm{I}}_{2M \times 2M}},\sum\limits_{i \in {\cal M}} {{{\cal I}^{\rm{T}}}{{\bm{\Phi }}_i}{\cal I}}  + \frac{1}{{\sigma _b^2}}} \right\}.
\end{align}
\end{subequations}

According to Schur complement, the EFIM of the position parameter ${\bm{J}}_{\rm{p}}^{\rm{e}}$ is given by
\begin{align}
{\bm{J}}_{\rm{p}}^{\rm{e}} = {\bm{A}} - {\bm{B}}{{\bm{C}}^{ - 1}}{{\bm{B}}^{\rm{T}}}.
\end{align}
Therefore, the EFIM in \eqref{Jpe} can be obtained.

\end{appendix}

\bibliographystyle{IEEE-unsorted}
\bibliography{refs1120}

\end{document}